# Observation of absorptive photon switching by quantum interference

Min Yan, Edward G. Rickey and Yifu Zhu

Department of Physics

Florida International University

Miami, Florida  33199

Abstract

We report an experimental demonstration of photon switching by quantum interference in a four-level atomic system proposed by Harris and Yamamoto (Phys. Rev. Lett. 81, 3611 (1998)). Quantum interference inhibits single-photon absorption but enhances third-order, two-photon type absorption in the four-level system. We have observed greatly enhanced nonlinear absorption in the four-level system realized with cold $^{87}$Rb atoms and demonstrated fast switching of the nonlinear absorption with a pulsed pump laser.

PACS numbers: 42.50.Gy, 42.50.Hz, 32.80.-t

Linear and nonlinear optical properties of an atomic medium can be modified by coherence and quantum interference that utilizes electromagnetically induced transparency (EIT) [1-8]. Because the vanishing linear absorption and large enhancement of nonlinear polarizability in an EIT medium, nonlinear optics may be studied at low light levels [9]. This has lead to interest of EIT related phenomena in multi-level atomic systems. There have been several studies of nonlinear spectral features in four-level atomic systems [10-12]. Recently, Harris and Yamamoto described a four-level EIT atomic system that exhibits greatly enhanced third-order susceptibility but has vanishing linear susceptibility [13]. In dispersive response, the EIT system exhibits giant Kerr nonlineaity [14] while in absorptive response, the EIT system absorbs only two photons, but not one photon. This feature may be used to realize an optical switch where a laser pulse controls absorption of another laser field at different frequency [13]. Recently, we reported an experimental study of the enhanced nonlinear absorption in the EIT system [15]. Here we report an experimental demonstration of absorptive photon switching in the four-level EIT system. We have observed switch on and off of the probe absorption for both a cw probe laser and a pulsed probe laser following the rise and fall of the switching laser pulse and the experimental results agree with theoretical calculations.

An energy-level diagram for $^{87}$Rb atoms used in our experiment is depicted in Fig. 1(a). A coupling laser drives the $D_1$ F=2→F'=1 transition at 795 nm and creates dressed atomic states |+> and |-> (|+>= $\frac{1}{\sqrt{2}}$ (|3>+|2>) and |->= $\frac{1}{\sqrt{2}}$ (|3>-|2>)). A weak probe laser drives the $D_1$ F=1→F'=1 transition and forms a standard Λ-type configuration for EIT. A switching laser drives the $D_2$ F=2→F'=3 transition at 780 nm. The probe laser and the switching laser are linearly polarized parallel with each other and perpendicular to the linearly polarized coupling laser. The induced transitions among the magnetic sub-levels by the three lasers can be grouped together according to the selection rules and form a manifold of four-level systems. To a good approximation, the coupled Rb system can be viewed as equivalent to a generic four-level system treated in Ref. [13]. The validity of such a simplification has been supported by several previous EIT-type studies in alkaline atoms [16-18]. For later discussions, g, Ω and Ω' are defined as Rabi frequencies for the probe transition |1>→|3>, the coupling transition |2>→|3>, and the switching transition |2>→|4> respectively. $\Delta_c=\omega_c-\omega_{32}$, $\Delta'=\omega_s-\omega_{42}$ and $\Delta=\omega_p-\omega_{31}$ denote the frequency detunings of the coupling laser, the switching laser and probe laser, respectively. The absorptive photon switching occurs under the condition, $\Delta_c=\Delta'=\Delta=0$.

Photon switching in the four-level atomic system is based on the interference enhanced nonlinear absorption and inhibited linear absorption. A simple physical picture can be given by a dressed state analysis [19]. It can be shown that the single-photon absorption experiences destructive interference and vanishes at the line center of the probe transition ($\Delta=0$) while the two-photon absorption in the dressed states interferes constructively, which leads to large enhancement at $\Delta=0$. If the dephasing $\gamma_{21}$ is negligible, the probe field only experiences nonlinear absorption at $\Delta=0$. The nonlinear absorption coefficient is given by $a(t) = K\,\mathrm{Im}(\rho_{13})$ ($K = \dfrac{N w_p d}{3V\varepsilon_0 \hbar c E_p}$). If the intensity of the input probe field is $I_{in}(t)$, then the intensity of the output probe field is given by $I_{out}(t) = I_{in}(t)e^{a(t)nl}$ (n is the atom density and $l$ is the absorption length). $\mathrm{Im}(\rho_{13})$ is negative and is proportional to the intensity of the switching laser [15]. So the probe absorption can be turned on and off by the switching laser, thus, realizing absorptive photon switching. $\mathrm{Im}(\rho_{13})$ is optimized at a moderate to weak coupling Rabi frequency $\Omega$ with the maximum value of $\mathrm{Im}(\rho_{13}) = -g/(2\Gamma_3)$ [15]. This value of the nonlinear absorption should be compared with the resonant linear absorption in a simple two-level system. In a simple two-level system (states |1> and |3> only) coupled by a weak laser field with a Rabi frequency g, the linear absorption amplitude at the resonance is given by $\mathrm{Im}(\rho_{13})=-g/\Gamma_3$. Therefore, the amplitude of the nonlinear absorption in the four-level EIT system may approach 50% of the amplitude of the linear absorption in an isolated two-level system [9]. Thus, the four-level system exhibits unusually large nonlinear absorption at low light intensities without any single-photon, linear absorption.

Our experiments is done in a vapor-cell magneto-optical trap (MOT) produced in the center of a 10-ports, 4-1/2" diameter, stainless-steel vacuum chamber pumped down to a pressure $\sim 10^{-9}$ torr. An extended-cavity diode laser with output power of $\sim$40 mW is used as the cooling and trapping laser. Another extended cavity diode laser with output power of $\sim$15 mW is used as the repump laser. The diameter of the trapping laser beams and the repumping laser beam is $\sim$ 1.5 cm. The trapped $^{87}$Rb atom cloud is $\sim$ 2 mm in diameter and contains $\sim 10^8$ atoms. The weak probe laser is provided by a third extended-cavity diode laser with a beam diameter $\sim$ 0.8 mm and power attenuated to $\sim$ 1 µW. The coupling field is provided by a fourth extended-cavity diode laser with a beam diameter $\sim$ 3 mm and output power ~ 20 mW. A Ti:Sapphire laser (Coherent 899-21) with a beam diameter ~ 3 mm is used as the switching laser. The simplified experimental scheme is depicted in Fig. 1(b). The probe laser, the coupling laser, and the switching laser are overlapped with the trapped Rb cloud and the transmitted light of the probe laser is recorded by a fast photodiode. The laser intensities are varied by neutral density filters. The experiment is run in a sequential mode with a repetition rate of 10 Hz and the time sequence is shown in Fig. 1(c). All lasers can be turned on or off by Acousto-Optic Modulators (AOM) according to the time sequence described below. For each period of $t_5-t_1$=100 ms, $t_5-t_4$=99.4 ms is used for cooling and trapping of the Rb atoms during which the trap laser and the repump laser are turned on by two separate AOMs and the coupling laser, the switching laser, and the probe laser are off. The time for the probe transmission measurements lasts ~ $t_4-t_1$=0.6 ms during which the trap laser and the repump laser are turned off and the coupling laser with its frequency fixed on the $D_1$ F=2→F'=1 transition is turned on by a third AOM. Three specific measurements are made according to the following configurations. First, for the measurement of the probe absorption spectrum, the weak probe laser is on all the time and the switching laser with its frequency fixed on the $D_2$ F=2→F'=3 transition is turned on by a fourth AOM at the same time $t_1$ as the coupling laser, and the pulse durations of the coupling laser and the switching laser are equal. After a 0.1 ms delay, the probe laser frequency is scanned across the $D_1$ F=1→F'=1 transition in ~0.1 ms and the probe transmission versus its frequency is recorded. Second, for the photon switching demonstration of the cw probe laser, the switching laser is turned on 0.1 ms after the coupling laser pulse and its pulse duration $t_3-t_2$ can be varied from 1 μs to 200 μs. The frequency of the probe laser is fixed on the $D_1$ F=1→F'=1 transition and the probe transmission versus time is recorded. Third, for the photon switching of a pulsed probe laser, the probe laser is turned on by a fifth AOM and has a pulse duration equal to that of the switching pulse as shown in Fig. 1(c). The transmitted probe pulse versus time is recorded.

Fig. 2 shows the measured transmission spectrum of the weak probe laser versus its frequency detuning Δ. The experimental data are plotted in solid lines while the dotted lines are numerical calculations of the four-level system depicted in Fig. 1(a). Fig. 2a shows the measured probe spectrum when the switching laser is on. The data are recorded under the condition of $\Delta'=\Delta_c$=0. The central peak (at Δ=0) in Fig. 2a corresponds to the pure two-photon type absorption enhanced by the constructive interference while the two side peaks at $\Delta \sim \pm \Omega/2$ represent the one-photon absorption of the dressed states. For comparison, Fig. 2b shows the measured probe spectrum under identical experimental conditions but without the switching laser. The nonlinear absorption peak at $\Delta=0$ disappears. It represents the EIT spectrum observed in the $\Lambda$-type Rb system. The coherence dephasing rate $\gamma_{21}$, due to the atom collisions and the off-resonant laser excitation, is estimated to be $\sim 10^3$ s$^{-1}$, which is negligibly small under our experimental conditions. Small residual absorption (~2%) at $\Delta=0$ in the EIT spectrum is likely due to the effect of the finite laser linewidths. From the EIT spectrum, the Rabi frequency of the coupling laser is deduced to be $\Omega \sim 9$ MHz. The Rabi frequencies of the switching laser and probe laser are estimated from the laser powers and beam diameters, which are $\Omega' \sim 15$ MHz, $g \sim 1$ MHz respectively. These parameters are then used in the numerical calculations shown by the dotted lines in Fig. 2. Since the laser pulse durations are much greater than the atomic decay times $1/\Gamma_3$ (30 ns) and $1/\Gamma_4$ (27 ns), the measurements are carried out essentially in the steady-state regime and the steady-state analysis is validated. Fig. 2a shows that the observed single-pass, probe nonlinear absorption at $\Delta=0$ is ~ 30%. From the measured percentage absorption of the two peaks in the EIT spectrum of Fig. 2b, we deduce that the linear probe absorption at the center of the $D_1$ F=1$\rightarrow$F'=1 transition without the coupling laser is ~ 70%. So the measured nonlinear absorption is ~ 42% of the linear absorption in a simple two-level system, which agrees with the theoretical calculation.

Fig. 3 shows the absorptive photon switching of the cw probe laser by a pulsed switching laser. The probe absorption is turned on and off by the switching pulse, which leads to a pulsed, stepwise probe transmission profile. The bandwidth of the two-photon absorption is determined by the EIT width that is given by the smaller of $\Omega$ and $\Omega^2/\Gamma_3$. Therefore, the switching speed is set by the EIT width [13]. Although a small EIT width is necessary for obtaining the maximum nonlinear absorption, it may compromise the optical switching time. As $\Omega$ decreases, the decay (fall) time of the probe absorption increases rapidly after the switching laser is turned off. The numerical calculation shows that for $\Omega=2\Gamma_3=10.6$ MHz, the decay time associated with the EIT width is $\sim 1/(2\pi\Gamma_3) \sim 30$ ns; for $\Omega=\Gamma_3=5.3$ MHz, the decay time increases to $\sim 2.5/(2\pi\Gamma_3) \sim 70$ ns. In our experiment, the rise and fall time of the switching laser pulse is ~ 150 ns (limited by the turn-on time of the AOM) so the optimal Rabi frequency that does not limit the switching time, yet still leads to a large nonlinear absorption, is $\Omega \sim 2\Gamma_3$. Then, the switching time of the probe absorption essentially follows the rise and fall time of the switching pulse. Fig. 3a plots the output intensity of the probe laser versus time for a pair of $\Omega$ values with a simulated switching pulse derived from the experiment. When $\Omega=2\Gamma_3$ (dashed curve), the switching time of the probe absorption follows the rise and fall time of the switching pulse; when $\Omega=\Gamma_3$ (solid curve), the decay time of the probe absorption becomes larger due to the smaller EIT width. The experimental data for the photon switching demonstration are plotted in Fig. 3b. The switching pulse with 1 µs duration (the upper curve) is turned on at t=0 (0.1 ms after the coupling laser is turned on). The probe transmission versus time (the two lower curves) shows that the probe absorption is switched on and off by the switching pulse and the peak absorption is about 30%. The observed switching time of the probe absorption in Fig. 3b follows the rise and fall time of the switching laser pulse for $\Omega$~10 MHz (dashed line). At $\Omega$~ 5 MHz (solid line), the decay time of the probe absorption becomes longer due to the smaller EIT width, in agreement with the theoretical calculations in Fig. 5a. We have varied the pulse duration of the switching laser from 1 µs to 200 µs and observed that the time duration of the probe absorption follows that of the switching laser pulse.

Fig. 4 shows the absorptive switching of a pulsed probe laser by a pulsed switching laser. The two pulses have the same duration (~1 µs) and overlap with each other in time as shown in Fig. 1(c). Fig. 4(a) plots the theoretical calculations of the normalized output intensity of the probe pulse $I_{out}(t)$. The rise (fall) time of the switching pulse and the input probe pulse are taken to be 150 ns to match the experimental conditions. The upper dashed line is the output probe pulse when the switching pulse is absent. The two lower traces are the output probe pulses when the switching pulse is on. The Rabi frequency of the coupling laser is $\Omega$~10 MHz (solid line) and 5 MHz (dot-dashed line). The output pulse of the probe laser shows a transient peak near the switching time t=0 in a time interval about equal to the lifetime of the excited state |3> and then decays to a quasi-steady-state value. The decay time depends on the EIT width and is longer for smaller $\Omega$ values. When the switching pulse and the input probe pulse are turned off, the output probe pulse is switched off smoothly and follows the fall time of the input pulse. Fig. 4b presents the measured output probe pulse without the switching laser pulse (the upper dashed line) and with the switching laser pulse (the two lower traces, corresponding to the $\Omega$ values in Fig. 4a). With the switching pulse on, the measured output pulses show the quick transient rise at the switching time and then exhibit the noticeable slower decay at the smaller $\Omega$ value due to the reduced EIT width, in agreement with the theoretical calculations. The intensity attenuation of the output probe pulse is about 30% and 25% for the two $\Omega$ values respectively. We have also varied the pulse duration of the switching laser and the probe laser from 1 µs to ~100 µs and observed that the transmitted probe pulse shows the transient rise and fall at the turn on time and then switches off smoothly with a lineshape similar to that shown in Fig. 4(b).

In conclusion, we have observed large third-order nonlinear absorption and demonstrated the absorptive photon switching by the constructive quantum interference in cold $^{87}$Rb atoms. Because the EIT cancellation of single-photon absorption, the nonlinear photon absorption is dramatically enhanced and the observed nonlinear absorption amplitude becomes comparable to that of the single-photon absorption in a two-level system. The experimental results can be interpreted in term of the two-photon absorption in the dressed state picture and the absorptive photon switch can therefore be viewed as a two-photon device. The experimental measurements agree with the theoretical calculations based on the four-level EIT system and provide an example of resonant nonlinear optics manifested by quantum interference under low light intensities.

This work is supported by the National Science Foundation and the U. S. Army Research Office.

Fig. 1 Yan et al.

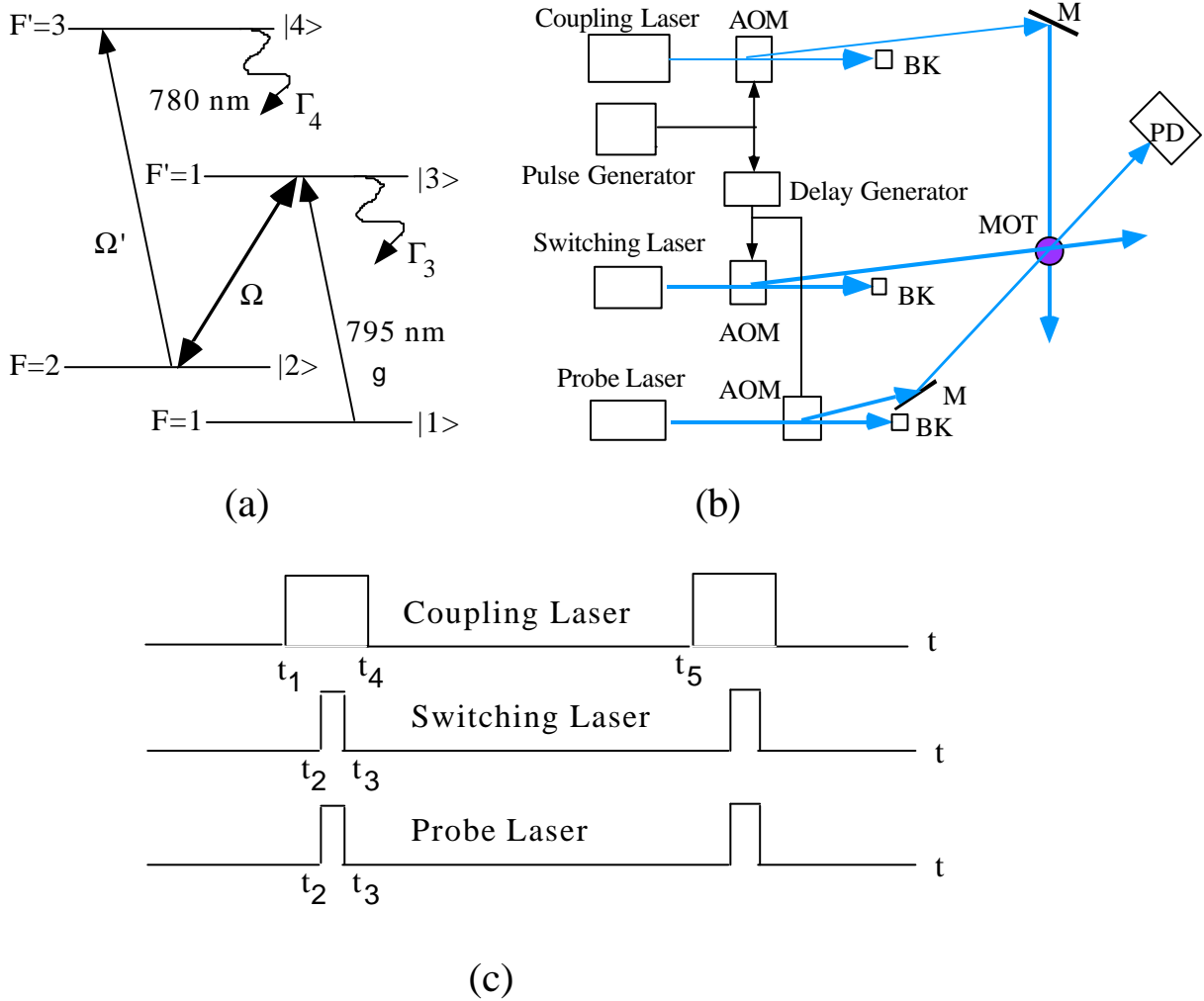

Fig. 1 (a) Four-level Rb atomic system and laser coupling scheme. $\Gamma_3$ ($2\pi \times 5.3$ MHz) and $\Gamma_4$ (($2\pi \times 5.9$ MHz) are the spontaneous decay rates. (b) Schematic diagram of the experimental set up. M: mirror; AOM: acousto-optic modulator; BK: beam blocker; PD: photodetector. (c) Experimental time sequence. $t_4-t_1=0.6$ ms, and the $t_5-t_4=99.4$ ms. The probe laser is shown to be configured for the pulsed switching. For the switching of a cw probe laser and the measurement of the probe nonlinear absorption spectrum, the probe laser is continuously on.

Fig. 2 Yan et al.

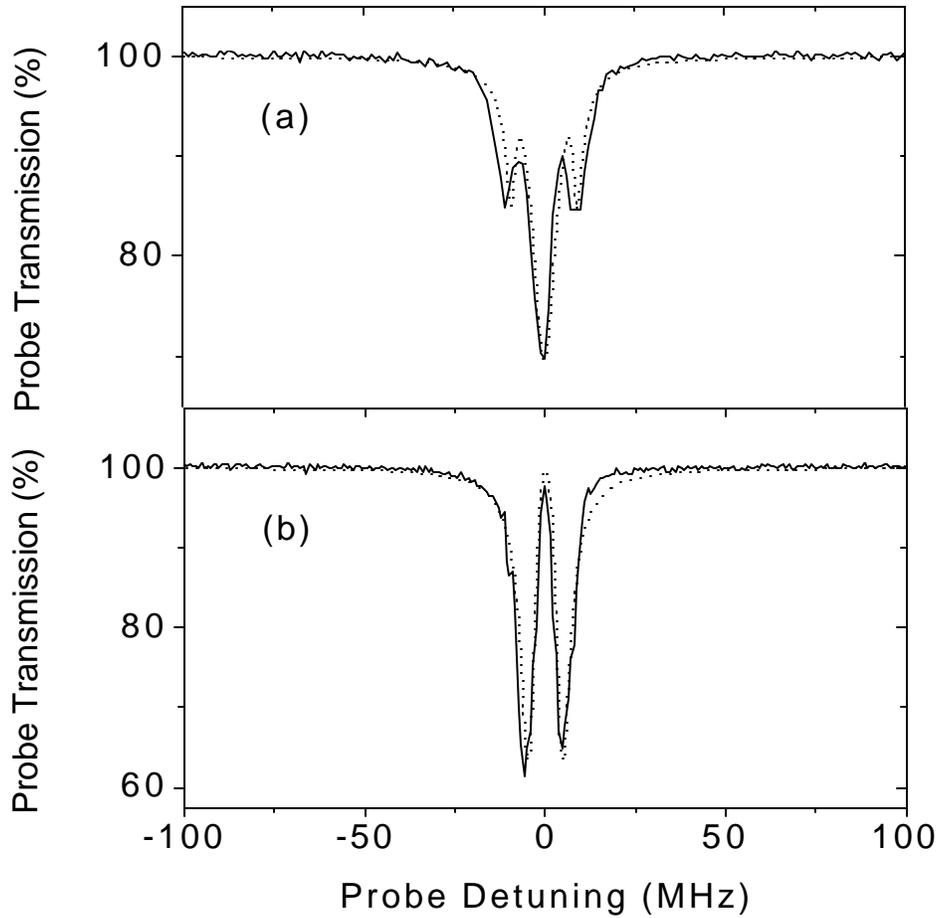

Fig. 2. (a) Measured probe transmission versus the probe frequency detuning $\Delta$ while the coupling laser and the switching laser are on resonance, i.e. $\Delta_c=\Delta'=0$. The central peak corresponds to the two-photon absorption and the two side peaks represent the Autler-Townes' doublet transitions. For comparison, the measured EIT spectrum of the probe transmission versus $\Delta$ is plotted in (b), which is taken without the switching laser. The solid lines represent the experimental data while the dotted lines are the theoretical calculations.

Fig. 3 Yan et al.

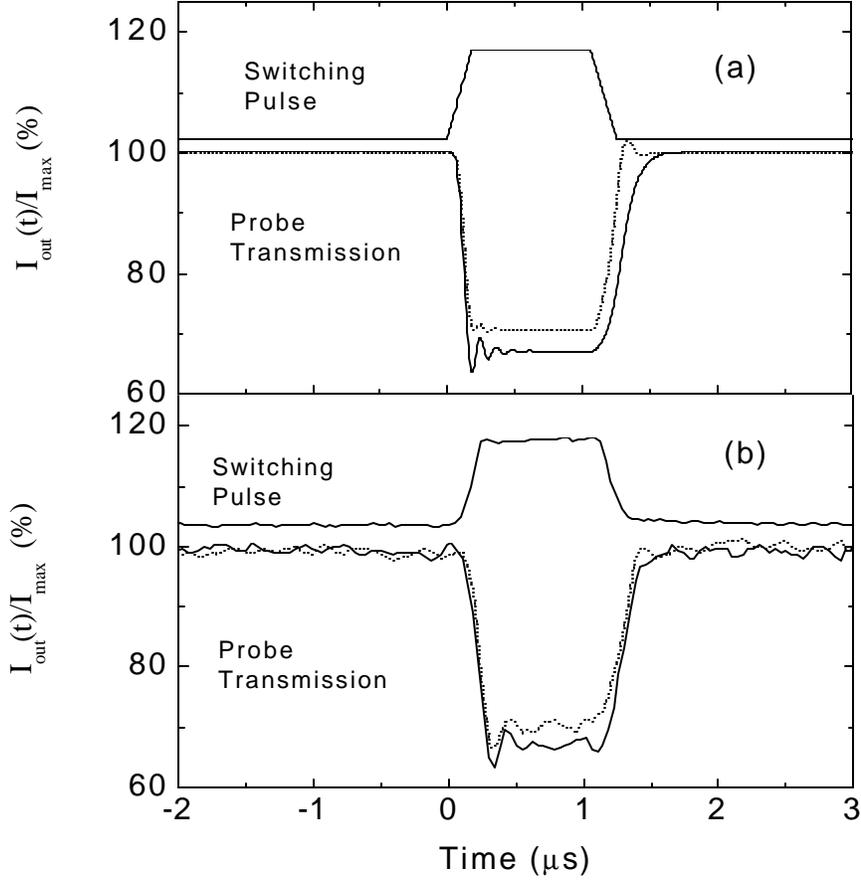

Fig. 3. (a) Calculated $I_{out}(t)/I_{max}$ of a cw probe laser versus time when a switching pulse simulated from the experiment (top curve) is turned on at $t=t_2=0$. $I_{max}$ is the amplitude of the input probe laser. The rise time of the switching pulse is 150 ns. The relevant parameters are $\Omega'=3\Gamma_3$, $g=0.2\Gamma_3$, and $\Omega=\Gamma_3$ (dotted line) and $2\Gamma_3$ (solid line) respectively. (b) The measured switching pulse (top curve in arbitrary units) and the probe transmission (lower curves) versus time. $\Omega\sim 10$ (5) MHz for the dotted (solid) curve.

Fig. 4   Yan et al.

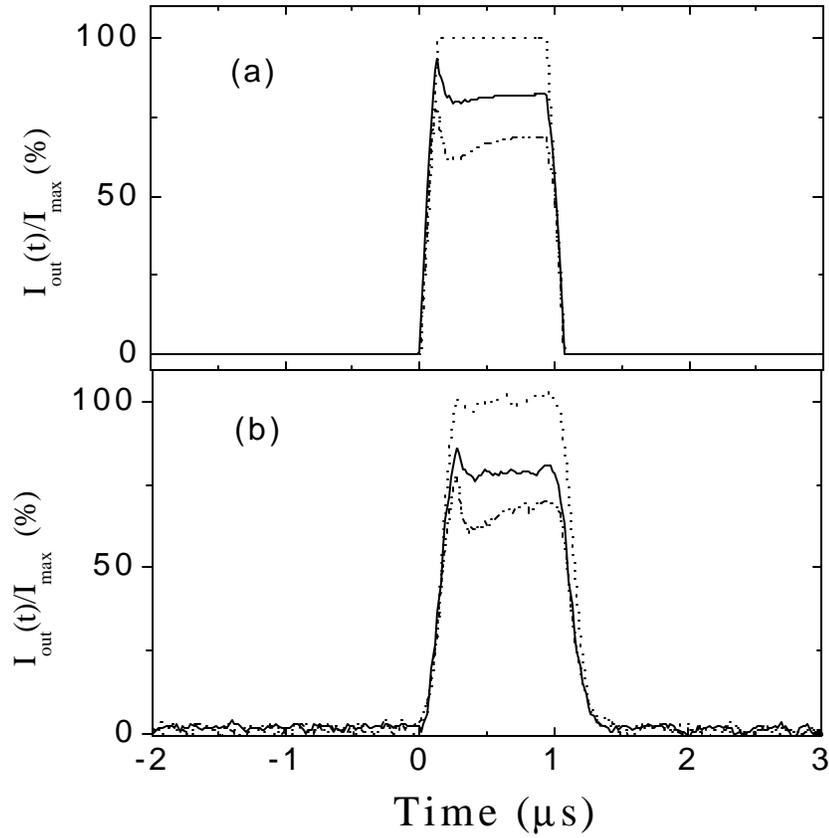

Fig. 4. (a) Calculated $I_{out}(t)/I_{max}$ of a pulsed probe laser versus time. The top dashed curve, $I_{out}(t)= I_{in}(t)$ (no switching pulse). $I_{max}$ is the amplitude of the input probe pulse. The two lower curves show the output probe pulse $I_{out}(t)$ when the switching pulse is turned on at $t=t_2=0$. The relevant parameters are $\Omega'=3\Gamma_3$, $g=0.2\Gamma_3$, and $\Omega=2\Gamma_3$ (solid line) and $\Gamma_3$ (dot-dashed line) respectively. (b) Measured output probe pulses, $I_{out}(t)$, versus time under the conditions given in (a).